\documentclass{emulateapj}
\usepackage{apjfonts}

\slugcomment{Accepted for publication in The Astrophysical Journal}
\shorttitle{WD HEATING AND COOLING IN DN OUTBURSTS}
\shortauthors{PIRO, ARRAS, \& BILDSTEN}

\newcommand{\be}{\begin{eqnarray}}
\newcommand{\ee}{\end{eqnarray}}
\newcommand{\beq}{\begin{equation}}
\newcommand{\eeq}{\end{equation}}
\def\tth{t_{\rm th}}
\def\tacc{t_{\rm acc}}
\def\tdn{t_{\rm dno}}
\def\delad{\nabla_{\rm ad}}
\def\del{\nabla}
\def\Teff{T_{\rm eff}}
\def\Tbl{T_{\rm bl}}
\def\lp{\left(}
\def\rp{\right)}

\begin{document}

\epsscale{1.2}

% -----------------------------------------------------------
% -----------------------------------------------------------

\title{White Dwarf Heating and Subsequent Cooling in Dwarf Nova Outbursts}

\author{Anthony L. Piro}
\affil{Department of Physics, Broida Hall, University of California,
	\\ Santa Barbara, CA 93106; piro@physics.ucsb.edu}

\author{Phil Arras}
\affil{NSF AAPF Fellow, Kavli Institute for Theoretical Physics, Kohn Hall, University of California,
	\\ Santa Barbara, CA 93106; arras@kitp.ucsb.edu}

\and

\author{Lars Bildsten}
\affil{Kavli Institute for Theoretical Physics and Department of Physics,
Kohn Hall, University of California,
	\\ Santa Barbara, CA 93106; bildsten@kitp.ucsb.edu}

% -----------------------------------------------------------
% -----------------------------------------------------------

\begin{abstract}
  We follow the time dependent thermal evolution of a white dwarf
(WD) undergoing sudden accretion in a dwarf nova outburst, using both
simulations and analytic estimates. The post-outburst
lightcurve clearly separates into early times when the WD flux is high,
and late times when the flux is near the quiescent level.  The break
between these two regimes, occurring at a time of order the outburst
duration, corresponds to a thermal diffusion wave reaching the base
of the freshly accreted layer. Our principal result is that long after
the outburst, the fractional flux perturbation about the quiescent flux 
decays as a power law with time (and {\it not} as an exponential). We use
this result to construct a simple fitting formula that yields estimates for both
the quiescent flux and the accreted column, i.e. the total accreted mass
divided by WD surface area. The WD mass is not well constrained by the
late time lightcurve alone, but it can be inferred if the accreted
mass is known from observations. We compare our work
with the well-studied outburst of WZ Sge, finding that the cooling is
well described by our model, giving an effective temperature
$T_{\rm eff}=14,500\ {\rm K}$ and accreted
column $\Delta y\approx10^6\ {\rm g\ cm^{-2}}$, in agreement with the
modeling of Godon et al. To reconcile this accreted column with the accreted
mass inferred from the bolometric accretion luminosity, a large WD mass
$\gtrsim1.1M_\odot$ is needed. Our power law result is a valuable
tool for making quick estimates of the outburst properties. We show that
fitting the late time lightcurve with this formula yields a predicted column
within $20\%$ of that estimated from our full numerical calculations.
\end{abstract}

% -----------------------------------------------------------
% -----------------------------------------------------------


\section{Introduction}

  Dwarf novae (DNe) are cataclysmic variables (CVs; Warner 1995) that
undergo dramatic accretion events in which a large fraction of the
accretion disk is dumped onto the white dwarf (WD) surface. These last
for $\sim2-20\ {\rm days}$ at accretion rates of
$\sim10^{-8} M_\odot\ {\rm yr}^{-1}$ and are separated by quiescent
intervals of $\sim10\textrm{ days}$
to tens of years. Following DN outbursts, CVs show a decreasing ultraviolet
flux \citep[e.g.,][]{lon94,gb96,sio96,szk98,che00}. The accretion rate then
is low enough that the WD surface can be directly seen,
and spectroscopic observations indicate this flux originates from a WD
cooling in response to the outburst \citep{ms84,kss91,lon93,sio93}. The
flux converges asymptotically to a quiescent level set by the time-averaged
accretion rate \citep{tb03}, $F_0=\sigma_{\rm SB}T_{\rm eff, 0}^4$, where
$\sigma_{\rm SB}$ is the Stefan-Boltzmann constant and $T_{\rm eff,0}$
is the quiescent WD effective temperature.

  The lightcurve of the cooling WD depends on the mechanism for heating,
which has previously been attributed to boundary layer irradiation
\citep{reg83,pri88,rs89,grs95,pop97}, compressional heating
\citep{sio95,gs02}, and shear
mixing luminosity \citep{spa93}. For a covering fraction $\la 0.1$
\citep{pb04}, we show in the Appendix that boundary layer heating 
(even via shear mixing) and surface advection have a negligible impact
on the late time cooling
lightcurve as long as the imposed surface temperatures are
$\lesssim10^6\ {\rm K}$. Temperatures higher than this are ruled out
according to theoretical work on the boundary layer \citep{pn95,pb04}.
We therefore focus solely on compressional heating to understand the
cooling flux released on long timescales. We show that this late time
cooling is a powerful tool for studying DN outbursts as it is independent
of the time dependent accretion rate during the outburst. Furthermore, WD
photospheric temperature estimates are less uncertain well after the
outburst since any residual screening material has dissipated
\citep[see][]{lon04}.

  We present numerical calculations following the compressional
heating and subsequent cooling of the WD surface in \S 2. We investigate
the late time cooling analytically in \S 3, showing that it obeys a power
law, and not an exponential decay
\citep[as has been assumed in other studies, for example in][]{gb96}.
Furthermore, this power law is not in temperature, but rather in the
{\it fractional temperature
perturbation about the quiescent WD temperature},
$\delta T_{\rm eff}/T_{\rm eff,0}$.  We construct a fitting formula
(eq. [\ref{eq:fit}]) that can be applied to DN cooling observations
to constrain both the quiescent flux, $F_0=\sigma_{\rm SB}T_{\rm eff,0}^4$,
and the column of material accreted during the DN outburst,
$\Delta y=\Delta M/(4\pi R^2)$ (using eq. [\ref{eq:deltay}]), where
$\Delta M$ is the total amount of mass accreted.
This late time lightcurve depends weakly on the surface WD gravity,
and therefore cannot directly constrain the WD mass. However, if $\Delta M$
is known from other observations, the WD radius (and mass) can
be inferred. We compare our model to the 2001 July outburst of WZ Sge in
\S 4, and show that its cooling is consistent with our model.
Our analytic formula fits the late time lightcurve and yields an accreted
column to within $20\%$ of that found from our full numerical calculation.
We conclude in \S 5 by summarizing our results and discussing the
importance of further, multi-epoch observations of WD cooling following
DN outbursts. An appendix investigates the impact of surface heating on
the late time cooling lightcurve.

% -----------------------------------------------------------
% -----------------------------------------------------------

\section{Time Evolution During the Dwarf Nova Outburst}

  In this section we present time dependent numerical models
of WD thermal evolution during and after a DN outburst.
These highlight the importance of deep compressional
heating, which will be useful for constructing the analytic late time
cooling curves.

% -----------------------------------------------------------

\subsection{Quiescent White Dwarf Model}

  We begin by describing the quiescent background model of the WD
envelope, as expected between outbursts when the accretion
rate is low ($\dot{M}\lesssim10^{-11} M_\odot\ {\rm yr}^{-1}$). We focus on
shallow, radiative surface layers of the WD, which we model with plane
parallel geometry and constant gravitational acceleration, $g=GM/R^2$. It
is convenient to use column depth $y=-\int \rho dr$ as a vertical
coordinate, where $r$ is the spherical radius. Hydrostatic balance
then becomes $P=gy$. Throughout this paper we give numerical
estimates using solar composition, ideal, nondegenerate gas and Kramer's
opacity. We made comparisons with a more detailed calculation using
OPAL opacities \citep{ir96} and did not find significant deviations
from our results. For solar composition and gravity $g=10^8\ {\rm cm\
s^{-2}}$, surface convection zones are confined to columns $y \la 1\
{\rm g\ cm^{-2}}$ for $T_{\rm eff} \gtrsim 11,000\ {\rm K}$, and can be 
ignored.

  In quiescence, the thermal profile is set
by the underlying flux from the core, determined by the time-averaged
accretion rate \citep{tb03}. The WD surface layers are described by the
radiative flux equation
\be
	F = \frac{16\sigma_{\rm SB} T^3}{3\kappa}
		\frac{\partial T}{\partial y},
	\label{eq:fluxeqn}
\ee
where $F$ is the outward directed flux, $\sigma_{SB}$ is the Stefan-Boltzmann
constant and $\kappa$ is the opacity, which we assume satisfies a power law
$\kappa=\kappa_0 \rho^a T^b$. For Kramer's opacity $a=1$, $b=-7/2$,
and $\kappa_0=6.5\times10^{22}$ (in cgs units). Denoting the quiescent
flux by $F_0=\sigma_{SB}T_{\rm eff,0}^4$, we integrate equation
(\ref{eq:fluxeqn}) to find the temperature profile in quiescence
\be
	T_0(y) & = & \left[ (n+1) \frac{F_0}{\alpha} \right]^{1/(4+a-b)}
			y^{1/(n+1)}
	\nonumber \\ & = &
	5.7\times 10^5\ {\rm K}\ g_8^{2/17} F_{0,12}^{2/17} y_5^{4/17},
	\label{eq:constflux}
\ee
where $g_8\equiv g/10^8\ {\rm cm\ s^{-2}}$,
$F_{0,12}\equiv F_0/10^{12}\ {\rm erg\ s^{-1}\ \rm cm^{-2}}$,
$y_5\equiv y/10^5\ {\rm g\ cm^{-2}}$,
$n=(3-b)/(1+a)$ is the polytrope
index ($n=13/4$ for Kramer's opacity), and
$\alpha\equiv(16\sigma_{SB}/3\kappa_0)(k_b/\mu m_p g)^a=6.2\times 10^{-27}g_8^{-1}$ (in cgs units).

  A property of the envelope that is important for understanding
the cooling is the local thermal time at a column depth $y$,
\be
	t_{\rm th, 0}(y) & = & \frac{y c_p T_0(y)}{F_0}
	\nonumber
	\\
	&=& \frac{c_p}{F_0}
		\left[ (n+1) \frac{F_0}{c_p} \right]^{1/(4+a-b)}
		y^{(n+2)/(n+1)}
	\nonumber \\ & = & 
	0.73\ {\rm yr}\ g_8^{2/17} F_{0,12}^{-15/17}y_5^{21/17},
	\label{eq:tth0}
\ee
where $c_p\approx5k_{\rm B}/(2\mu m_p)$ is the specific heat,
$k_{\rm B}$ is Boltzmann's constant, and $\mu=0.62$ is the mean molecular
weight in the nondegenerate plasma of solar composition.

% -----------------------------------------------------------

\subsection{Compressional Heating During the Outburst}
\label{sec:heating}

  The entropy equation describing time evolution of the
WD temperature profile during the DN outburst can be written as \citep{bil98}
\be
	\frac{\partial F}{\partial y}
		= c_p \left[ \frac{\partial T}{\partial t}
		+ \frac{\dot{m}T}{y} \left( \nabla-\nabla_{\rm ad} \right)
			\right],
	\label{eq:entropy}
\ee
where $\dot{m}\equiv\dot{M}/(4\pi R^2)$ is the accretion rate per unit
area, $\nabla\equiv\partial\ln T/\partial\ln y$ is found from equation
(\ref{eq:fluxeqn}), and $\nabla_{\rm ad}\equiv (\partial\ln T/\partial\ln
P)_{\rm ad}\approx 2/5$ sets the adiabatic profile. The second term on the
right hand side arises from the advection of  entropy $v_r \partial s/\partial
r = (-\dot{m}/\rho) \partial s/\partial r = \dot{m}\partial s/\partial
y$ by the accretion flow, where $v_r$ is the advection velocity and $s$
is the specific entropy.

  We solve equations (\ref{eq:fluxeqn}) and (\ref{eq:entropy}) implicitly
in time using backward time differencing for stability and typically
128 to 512 grid points. The time step is chosen so that the average
temperature at each grid point changes by a fractional
amount $\lesssim10^{-3}$ from one time step to the next. We
repeated a number of runs with a fractional change of
$10^{-4}$ and the results did not change.
The boundary condition at the base of the layer is taken to
be constant flux, $F_0$, and set at a depth ($\approx10^8\ {\rm g\ cm^{-2}}$)
where the local thermal time is longer than any timescale of interest
(i.e. longer than the time we wish to follow the cooling).
The flux into the core does change
due to each DN outburst, but by a tiny amount of order the ratio of the
mass accreted in one DN outburst to the envelope mass.
As we do not model the photosphere, the
surface boundary condition is set at a shallow depth with a sufficiently
short thermal time for the temperature profile to be constant flux,
$T \propto y^{1/(n+1)}$. The initial condition for the outburst is the
constant flux profile in equation (\ref{eq:constflux}). This is evolved
forward for a time $\tdn$, with mass accretion set by a given function
$\dot{m}(t)$.

\begin{figure}[tbp]
\plotone{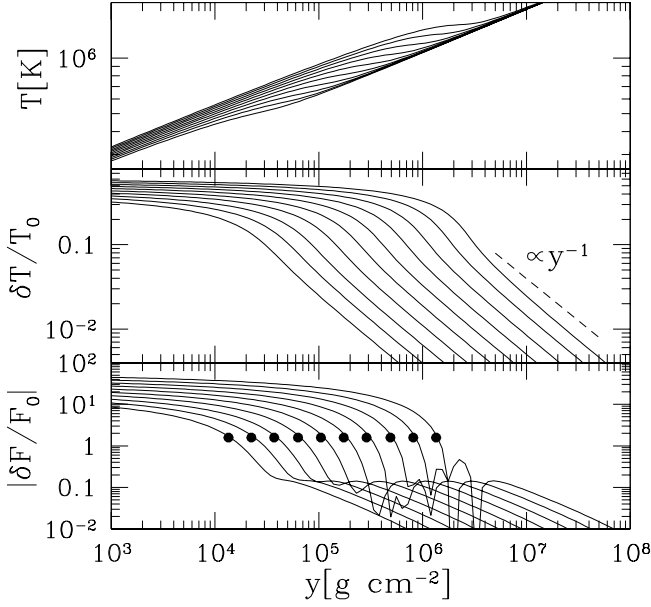}
\caption{Temperature, $T$, fractional temperature perturbation,
$\delta T/T_0$, and fractional flux perturbation, $|\delta F/F_0|$, during
accretion with $\dot{m}=0.30\ {\rm g\ cm^{-2}\ s^{-1}}$,
$\tdn=52\ {\rm days}$, and $T_{\rm eff,0} =14,500\ {\rm K}$. The 10
curves are logarithmically spaced in time from $0.52$ to $52\ {\rm days}$,
from bottom to top. The dots in the bottom panel mark the base of the
accreted layer, $\dot{m}t$. At some depths
$F(y,t)<F_0$, so that $\delta F/F_0<0$. Since we plot the quantities
logarithmically to emphasize the power laws, we must take the absolute
value of $\delta F/F_0$, which creates the jumps seen in the bottom
panel whenever $\delta F/F_0$ passes through zero.}
\label{fig:heatingprofile}
\end{figure}

  In Figure \ref{fig:heatingprofile} we plot the heating profiles during a
DN outburst at 10 different times, spaced logarithmically from 0.52
to 52 days. The WD has
$T_{\rm eff, 0}=14,500\ {\rm K}$ and $g=10^8\ {\rm cm\ s^{-2}}$
(as appropriate for an $M=0.6M_\odot$ WD), and accretes at a constant
rate $\dot{m}=0.30\ {\rm g\ cm^{-2}\ s^{-1}}$ [a global accretion rate of
$6\times10^{-8}M_\odot\ {\rm yr^{-1}}(R/10^9\ {\rm cm})^2$], giving a
total accreted column $\Delta y=1.3\times10^6\ {\rm g\ cm^{-2}}$. We
focus on the fractional temperature and flux profiles,
$\delta T(y,t)/T_0(y)= (T(y,t)-T_0(y))/T_0(y)$ and
$\delta F(y,t)/F_0=(F(y,t)-F_0)/F_0$, respectively, as we later show
that these quantities obey power laws at late times.

  The bottom panel in Figure \ref{fig:heatingprofile} shows the flux
perturbation as a function of depth, increasing in time from bottom
to top. Two distinct regimes are seen for $\delta F/F_0$, separated
by the base of the accreted layer, shown by dots. Above
this depth, the profiles are determined by the steady state solutions
to equation (\ref{eq:entropy}), while below the perturbations are small
and largely due to adiabatic compression. The middle panel shows the
fractional temperature perturbation.  Near the surface, this quantity
is fairly constant with a slight decrease with depth, but below the base
of the accreted layer it falls off as a power law $\delta T/T_0 \propto
y^{-1}$ for adiabatic compression.  Finally, the absolute temperature
is shown in the top panel. The flattening of the temperature profile
near the  base of the accreted layer is apparent, especially at late
times. If the outburst had lasted longer, or the accretion rate was
higher, a temperature inversion would have developed just below the base
of the accreted layer. Since $t_{\rm th}$ is short in these shallow layers,
such an inversion ``smooths out'' very quickly once the envelope begins
to cool and has a negligible effect on the cooling lightcurve.

  The final profile after the DN event can be understood by comparing the
timescales of equation (\ref{eq:entropy}), which are the accretion time
$\tacc = y/\dot{m}$, the thermal time $t_{\rm th}=yc_pT/F$, and the elapsed
time $t$. Above the base of the accreted layer, $y\ll\dot{m}t$, there is a
hierarchy of timescales $t_{\rm th} \ll t_{\rm acc} \ll t$, so that
we take $\partial/\partial t\approx0$.  The flux near the surface is
nearly constant, with a small decrease with depth due to compressional
heating. Denoting the (constant) surface flux as $F_s$, we find the
temperature profile associated with this, which we denote $T_s(y)$,
by using equation (\ref{eq:constflux}) with $F_0$ replaced by $F_s$.
Substituting $T_s(y)$ into equation (\ref{eq:entropy}) and integrating
we find
\be
	F(y) & \approx & F_s - \left( \frac{\nabla_{\rm ad}}{\nabla}-1 \right)
		\dot{m} c_p T_s(y).
	\label{eq:surfaceflux}
\ee
This shows how the flux decreases near the surface, as can be seen in Figure
\ref{fig:heatingprofile}. The true temperature profile is therefore smaller
than the constant flux profile, $T_s(y)$, and can be estimated by substituting equation
(\ref{eq:surfaceflux}) into equation (\ref{eq:fluxeqn}) and integrating,
\be
	T(y) & \approx & T_s(y) \left[ 1 
	- \frac{1}{5+a-b}\left( \frac{\nabla_{\rm ad}}{\nabla}-1 \right) 
		\frac{\dot{m} c_p T_s(y)}{F_s} \right].
\ee
Setting $F \sim 0$ at the base of the accreted
layer, $y\approx \dot{m}t_{\rm dno}$
(see Figure \ref{fig:heatingprofile}), we can estimate the flux $F_s$ as a function
of $t_{\rm dno}$. This implies an effective temperature during accretion
\be
	T_{\rm eff} &=&  \left( \frac{F_s}{\sigma_{SB}} \right)^{1/4}
	\nonumber
	\\
	&\approx& 3.0\times 10^4\ {\rm K}\ g_8^{1/30}
	\nonumber
	\\
	&&\times
	\lp\frac{\dot{m}}{0.3\ {\rm g\ cm^{-2}\ s^{-1}}}\rp^{7/20}
	\lp\frac{t_{\rm dno}}{1\ {\rm day}}\rp^{1/15},
\ee
which agrees with our numerical models. This power law index of
$1/15\approx0.067$ is similar to the power laws found by \citet{gs03}
during heating. We also agree
that the power law during heating is for the actual
effective temperature (as opposed to the {\it fractional} effective
temperature as we find for the cooling). Since this evolution takes
place coincident with a high accretion rate, it will not be
observable so we do not elaborate on its effects any more here.

  We now derive the profile for $\delta T/T_0$ deep in the envelope due
to adiabatic compression. This is the first step in understanding
late time cooling. Well below the base of the accreted layer,
$y\gg\Delta y$, the timescales satisfy $t \ll \tacc \ll \tth$.
The initial temperature profile from equation
(\ref{eq:constflux}) is compressed adiabatically, giving
\be
	T(y) & = & T_0(y-\Delta y) \left(\frac{y}{y-\Delta y} \right)^{\delad}
	\nonumber \\ & \approx & 
	T_0(y) \left[ 1 + \left( \delad - \del \right) \frac{\Delta y}{y}
		\right],
\ee
or, a fractional temperature perturbation
\be
	\frac{\delta T}{T_0} & \approx & \left( \delad - \del \right)
		\frac{\Delta y}{y},
	\label{eq:deltat}
\ee
which explains the scaling $\delta T/T_0\propto y^{-1}$ in
Figure \ref{fig:heatingprofile}.
This extends from deep in the
envelope up to $y \approx \Delta y = \dot{m}\tdn$.

% -----------------------------------------------------------

\subsection{Cooling Following the Outburst}
\label{sec:cooling}

\begin{figure}[tbp]
\plotone{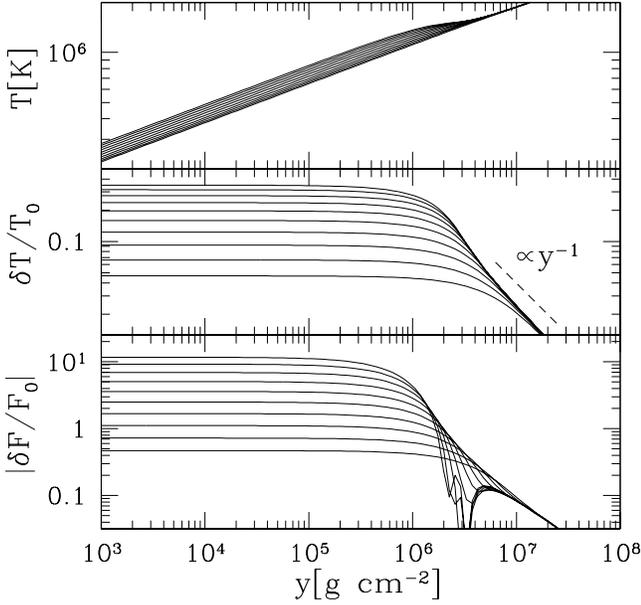}
\caption{Profiles for $T$, $\delta T/T_0$, and $|\delta F/F_0|$ during the
cooling following the outburst presented in Figure \ref{fig:heatingprofile},
in which a column $\Delta y=1.3\times 10^6\ {\rm g\ cm^{-2}}$ was accreted
in $\tdn=52\ {\rm days}$. The 10 curves are logarithmically spaced in time
from $10$ to $1000\ {\rm days}$, from top to bottom.}
\label{fig:coolingprofile}
\end{figure}

 We evolve the final profile from Figure \ref{fig:heatingprofile}
forward in time
to see how the WD cools after accretion has halted. Again we solve
equations (\ref{eq:fluxeqn})
and (\ref{eq:entropy}), but now with $\dot{m}=0$. In Figure
\ref{fig:coolingprofile} we show profiles during cooling spaced
logarithmically in time from 10 to 1000 days after the end of the outburst. Near the
surface, the profiles show constant perturbations $\delta F/F_0$
and $\delta T/T_0$, while at large depths the profile is unevolved.
Just as during heating, $\delta F/F_0$ is an order of magnitude larger than
$\delta T/T_0$, which we explain  in \S \ref{sec:selfsimilar} when
we derive self-similar solutions for the cooling phase.

  From this example we see that cooling involves a thermal wave moving
into the envelope, propagating down to a column $y$ only at a time,
$t\approx\tth(y)$. This is simplest to show for the deep, adiabatically
compressed material. Since the perturbations at these depths are small,
we approximate $\tth\approx t_{\rm th, 0}$. Setting $t_{\rm th,0}=t$ we invert
equation (\ref{eq:tth0}) to find the column that is just beginning to cool as a
function of time,
\be
	y_{\rm th}(t) &=& \left(\frac{F_0}{c_p}\right)^{(n+1)/(n+2)}
		\left[ \frac{\alpha}{(n+1)F_0} \right]^{1/(n+2)}
		t^{(n+1)/(n+2)}
	\nonumber \\ & \approx & 
	1.3\times 10^5\ {\rm g\ cm^{-2} }\ g_8^{-2/21} F_{0,12}^{15/21} 
	\lp\frac{t}{1\ {\rm yr}}\rp^{17/21},
	\label{eq:yth}
\ee
where $t$ is measured from the end of the outburst.
Substituting this result for $y$ in equation (\ref{eq:deltat}), we find
that the late time cooling should evolve according to
$\delta T/T_0\sim(\nabla_{\rm ad}-\nabla)\Delta y/y_{\rm th}(t)\propto t^{-(n+1)/(n+2)}$.
As we shall see in \S \ref{sec:selfsimilar}, this estimate is correct up
to a numerical  factor that can be computed.

  In the Appendix we consider the effects of boundary layer heating on
this late time cooling. We find that surface heating only propagates a
finite distance into the envelope, while compressional heating causes
a $\delta T/T_0\propto y^{-1}$ perturbation down to arbitrarily large
depths, so that at very late times compressional heating always
dominates. In practice, this may occur so long after the outburst that
during the observational period both heating mechanisms must be
considered for proper modeling. This happens when the boundary
layer is too hot, so that a diffusion wave propagating from the surface
does not have sufficient time to reach the background quiescent profile
during heating, resulting in a critical boundary layer temperature
(eq. [\ref{eq:tblcrit}]),
\be
	T_{\rm bl,crit}\sim10^6\ {\rm K}\ g_8^{2/17}F_{0,12}^{2/17}
	\lp\frac{\Delta y}{10^6\ {\rm g\ cm^{-2}}}\rp^{4/17}.
\ee
This critical temperature is close to, but most likely ruled out by
observations \citep{mau04} and theoretical modeling \citep{pn95,pb04}.

% -----------------------------------------------------------
% -----------------------------------------------------------

\section{Self-Similar Cooling}
\label{sec:selfsimilar}

  In \S \ref{sec:cooling}, we found that the fractional temperature
perturbation decreases as a simple power law with time for late time
cooling after an outburst. We now provide a rigorous
derivation of the power law index and calculate the prefactor.

  The discussion in \S 2.3 suggests that
the late time cooling is governed by the self-similarity variable
\be
	\xi(y,t) & = & \left[ \frac{t_{\rm th,0}(y)}{t} \right]^{1/2}
	= \left[ \frac{y}{y_{\rm th}(t)} \right]^{(n+2)/2(n+1)},
\ee
where $t$ is measured from the end of the outburst, and
we have introduced the $1/2$-power for calculational
convenience. At large depths, the initial temperature
profile is given by equation (\ref{eq:deltat}), so we make
the ansatz 
\be
	\delta T(y,t) & = & T_0(y) \left( \nabla_{\rm ad}-\nabla \right)
	\frac{\Delta y}{y} f(\xi),
	\label{eq:deltatselfsim}
\ee
where $f(\xi)$ is an undetermined function that contains the time
dependence of the temperature perturbation. To understand
the late time cooling we now find $f$ for large $t$
(i.e. in the limit $\xi\rightarrow0$).

  Taking perturbations of equation (\ref{eq:entropy}) about the
quiescent profile (with $\dot{m}=0$) gives
\be
	c_p \frac{\partial \delta T}{\partial t} & = &
		\frac{\partial \delta F}{\partial y},
	\label{eq:perturbedentropy}
\ee
where the perturbed flux is
\be
	\frac{\delta F}{F_0} & = & 
	(n+1) \frac{\partial}{\partial \ln y} \left( \frac{\delta T}{T_0}
			\right)
		+ (4+a-b) \frac{\delta T}{T_0}.
	\label{eq:perturbedflux}
\ee
We substitute the perturbed temperature from equation
(\ref{eq:deltatselfsim}) into equations (\ref{eq:perturbedentropy})
and (\ref{eq:perturbedflux}), which are then combined to give a single
equation for $f$,
\be
	\xi \frac{df}{d\xi} \left[ \frac{\xi^2}{2} - (n+2)
	+ \frac{(n+2)^2}{4(n+1)}
	+ \frac{(n+2)}{2(n+1)}(4+a-b) \right]
	\nonumber
	\\
	+\frac{(n+2)^2}{4(n+1)} \xi^2 \frac{d^2f}{d\xi^2} 
		+ f \left[ n+1 - (4+a-b) \right]  =  0.\hspace{0.5cm}
	\label{eq:selfsimeqn}
\ee
In the limit $\xi \rightarrow \infty$, $f \rightarrow 1$ by definition. In the
other limit of $\xi \rightarrow 0$, equation (\ref{eq:selfsimeqn}) has
a finite solution $f = \eta\xi^{2(n+1)/(n+2)}$, where $\eta$ is a constant
of proportionality determined by the full numerical solution.
Since equation (\ref{eq:selfsimeqn}) contains unwanted solutions that
are strongly divergent, we solve it as a matrix equation, finding
$\eta= 0.141$ for Kramer's opacity. The late
time temperature perturbation is then
\be
	\frac{\delta T(y,t)}{T_0(y)}
	& = & \eta\left( \nabla_{\rm ad}-\nabla \right)
	\frac{\Delta y}{y_{\rm th}(t)},
\ee
which is what we guessed in \S 2.3, up to the factor $\eta$.

  To make comparisons with the observed late time cooling, we
relate the above analysis to the fractional $T_{\rm eff}$ perturbation,
$\delta T_{\rm eff}/T_{\rm eff,0}=(T_{\rm eff}-T_{\rm eff,0})/T_{\rm eff,0}
\approx\delta F/4F_0$. At late times, 
\be
	\frac{\delta T}{T_0} & \propto & y^{-1} \xi^{2(n+1)/(n+2)}
	\approx \mbox{constant in $y$}.
\ee
Equation (\ref{eq:perturbedflux}) then relates the temperature and flux
perturbations
\be
	\frac{\delta F}{F_0}
	& \approx & \left( 4+a-b \right) \frac{\delta T}{T_0},
\ee
where the factor $4+a-b=8.5$ explains the order of magnitude difference
between $\delta F/F_0$ and $\delta T/T_0$.
Both $\delta F/F_0$ and $\delta T/T_0$ are constant in space in the limit
$t \gg t_{\rm th}(y)$.
We put these results together to express
the change of $\delta T_{\rm eff}$ in the linear limit in the form
\be
	\frac{\delta T_{\rm eff}}{T_{\rm eff,0}} & \approx &
	\left( \frac{t_{\rm late}}{t} \right)^{(n+1)/(n+2)}
	\label{eq:fit},
\ee
where $t$ is the time since the outburst ended,
the power law index is $(n+1)/(n+2)=17/21\approx0.81$, and
the characteristic late time cooling timescale is
\be
	t_{\rm late} & = & \left[ \eta \frac{4+a-b}{4} 
	\left( \nabla_{\rm ad} - \nabla \right) 
	\right]^{(n+2)/(n+1)} 
	t_{\rm th,0}(\Delta y)
	\nonumber
	\\
	&\approx& 0.024\ t_{\rm th,0}(\Delta y).
\label{eq:tlate}
\ee
Observations that fit $t_{\rm late}$ then constrain
\be
	\Delta y &=& 1.8\times10^4\ {\rm g\ cm}^{-2}
	\nonumber
	\\
	&&\times g_8^{-2/21}\lp\frac{t_{\rm late}}{1\ {\rm day}}\rp^{17/21}
		\lp\frac{T_{\rm eff,0}}{10^4\ {\rm K}}\rp^{60/21},
	\label{eq:deltay}
\ee
the total accreted column during the outburst.

\begin{figure}[tbp]
\plotone{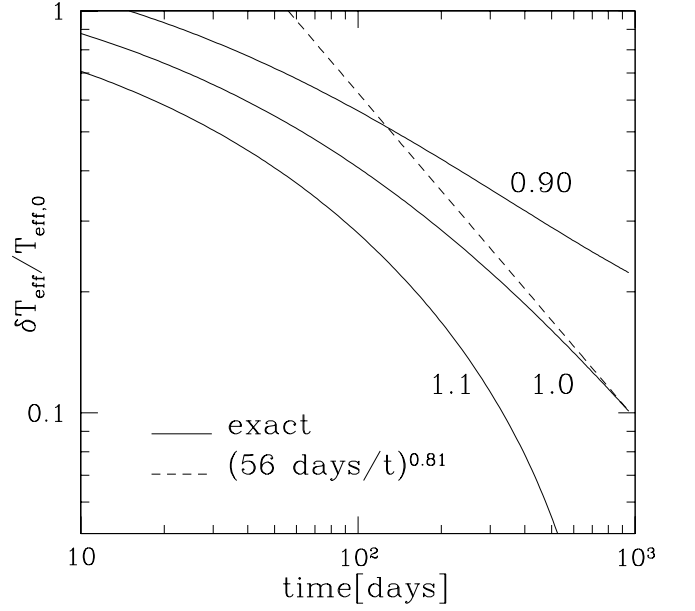}
\caption{The $T_{\rm eff}$ perturbation during cooling from the profiles
shown in Figure
\ref{fig:coolingprofile}. The three solid lines labeled $x=0.9,1.0,1.1$
represent $(T_{\rm eff}(t)-xT_{\rm eff,0})/(xT_{\rm eff,0})$, to show how
a power law with index $\approx0.81$ only results if the correct
$T_{\rm eff,0}$ is used. The analytic formula given in equation (\ref{eq:fit})
is shown by the dashed line. } \label{fig:latetimecomp}
\end{figure}

  Equations (\ref{eq:fit}) and (\ref{eq:deltay}) are our central results.
We make the following remarks on their use for understanding
observations:
\begin{itemize}
\item Observations should be plotted as $\delta T_{\rm eff}/T_{\rm eff,0}$ in
order to test whether a power-law is observed.  Equation (\ref{eq:fit})
has three parameters, $T_{\rm eff,0}$, $t_{\rm late}$, and
$n$, the polytrope index. If the wrong $T_{\rm eff,0}$ is used,
the late time cooling does not look like a power law with index
$\approx0.81$, giving a useful constraint on $T_{\rm eff,0}$.
Figure \ref{fig:latetimecomp} shows the result for $(T_{\rm
eff}(t)-xT_{\rm eff,0})/xT_{\rm eff,0}$ from the full cooling code,
where $x=0.9,1.0,1.1$. At late times, the analytic
formula given by equations (\ref{eq:fit}) and (\ref{eq:deltay})
agrees well. When the wrong quiescent effective temperature is used
the lightcurve either does not look like a power law
(as in the case of too large of a $T_{\rm eff,0}$, i.e. $x=1.1$), or
shows a power law that differs considerably from a power law index
of 0.81 (as in the case of too small of a $T_{\rm eff,0}$, i.e. $x=0.9$).
\item If the temperature perturbation $\delta\Teff/T_{\rm eff,0}\ga0.1$,
effects second order in $\delta\Teff/T_{\rm eff,0}$ cause the lightcurve
to deviate slightly
from a power law, as is apparent in Figure \ref{fig:latetimecomp} for
$x=1.0$ and $t\lesssim400\ {\rm days}$. However, adjusting $T_{\rm eff,0}$
until the data is nearly a power law with index 0.81 and applying the
analytic formula gives quick results accurate to  $\sim 20\%$. We discuss
this point in more detail in \S \ref{sec:wzsge}.
\item Given a measurement of $t_{\rm late}$, we can constrain $\Delta
y g^{2/21}$ using equation (\ref{eq:deltay}).  If we allow WDs
in the mass range $0.6-1.2M_\odot$, the factor $g^{2/21}$ can vary by
$\sim6^{2/21}\approx 1.2$. Therefore, if the WD mass is not known, we
can constrain $\Delta y$, the accreted {\it column} to 20\%!
\end{itemize}

\begin{figure}[tbp]
\plotone{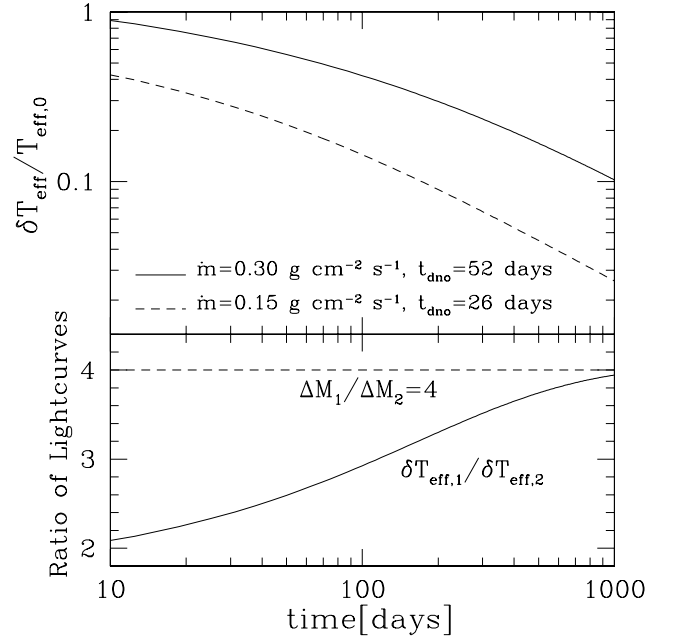}
\caption{A comparison of two outbursts, one accreting
at a rate of $0.30\ {\rm g\ cm^{-2}\ s^{-1}}$ for $\tdn=52\ {\rm days}$ and the
other at a rate of $0.15\ {\rm g\ cm^{-2}\ s^{-1}}$ for $\tdn=26\ {\rm days}$,
but both with $T_{\rm eff,0}=14,500\ {\rm K}$.
This results in four times the mass being accreted in the former outburst in
comparison with the latter.
In the top panel we show the two resulting $\delta T_{\rm eff}/T_{\rm eff,0}$
lightcurves, while in the bottom panel we plot the ratio of these two
lightcurves. This shows that at late times the ratio asymptotes toward
a value equal to the ratio of total accreted masses. A similar study would be
useful when considering the multiple outbursts from a given CV.}
\label{fig:mass}
\end{figure}

  If multiple outbursts are observed from the same system, the lightcurves
can be compared to infer the relative amount of mass accreted. We combine
equations (\ref{eq:fit}) and (\ref{eq:deltay}) to find
\be
	\frac{\Delta M_1}{\Delta M_2}
	\approx\frac{\delta T_{\rm eff,1}(t)}{\delta T_{\rm eff,2}(t)},
	\label{eq:mass}
\ee
where $\Delta M$ is the amount of mass accreted in an outburst and the
subscripts refer to the two outbursts. In this way disk instability models
can be tested, including understanding the differences between normal DN
outbursts and superoutbursts. In Figure \ref{fig:mass} we show how this
could be done by comparing the lightcurves of two outbursts, one accreting
at a rate of $0.30\ {\rm g\ cm^{-2}\ s^{-1}}$ for $\tdn=52\ {\rm days}$ and the
other at a rate of $0.15\ {\rm g\ cm^{-2}\ s^{-1}}$ for $\tdn=26\ {\rm days}$,
but both with $T_{\rm eff,0}=14,500\ {\rm K}$, so that four times the mass is
accreted in one outburst in comparison with the other. In the top panel we
compare the two resulting lightcurves, and in the bottom panel we look at
the ratio of these two lightcurves, which shows how this ratio asymptotes
to the accreted mass ratio at late times. For this to work correctly, the two
lightcurves must be correctly positioned in time, with $t=0$ corresponding
to the end of the outburst. Fortunately, at late times (when this works
best) it should not be difficult to do this with enough accuracy (within
$\sim10\ {\rm days}$) for useful results.

% -----------------------------------------------------------
% -----------------------------------------------------------

\section{The 2001 July Outburst of WZ Sge}
\label{sec:wzsge}

  The most detailed observation of a decay after a DN outburst is from the
2001 July 23 outburst of WZ Sge \citep{ish01}, providing an excellent test
of our cooling model. The outburst lightcurve is complicated \citep{pat02},
initially exhibiting a period of high $\dot{M}$ for 24 days (with a steady
decline), a sharp drop in $\dot{M}$ for 3 days, and finally quasi-periodic
accretion from days 29 to 52.

\begin{figure}[tbp]
\plotone{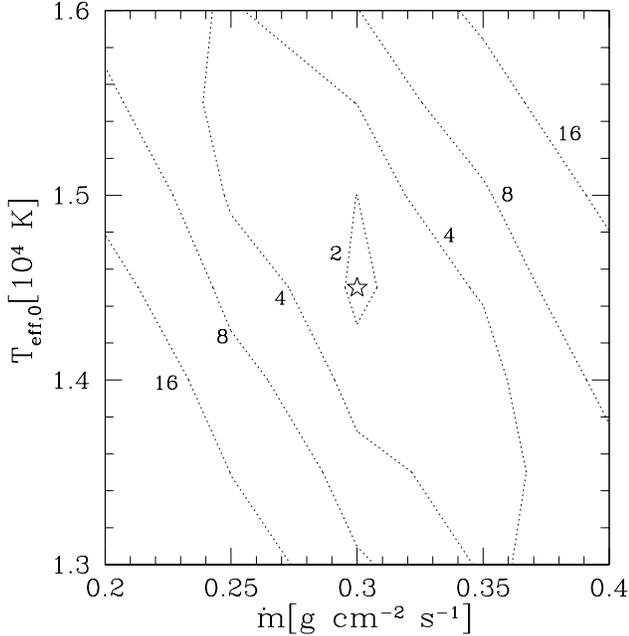}
\caption{Contours of constant $\chi^2$ (dashed lines) from fitting the
eight temperature measurements during the cooling of the 2001 July
outburst of WZ Sge. We vary both $\dot{m}$ and $T_{\rm eff,0}$ over a
grid of our numerical models, but keep
$t_{\rm dno}=52\ {\rm days}$ fixed. The star marks
our best fit model, $\dot{m}=0.30\ {\rm g\ cm^{-2}\ s^{-1}}$ with
$T_{\rm eff,0}=14,500\ {\rm K}$.}
\label{fig:contour}
\end{figure}

  The surface temperature of the WD during cooling has been estimated
three different ways \citep{lon04,god04}. For the sake of comparison, we
fit the median temperature measurement \citep[denoted ``$T_b$'' in][]{god04},
and assume an error $\approx1000\ {\rm K}$ (G\"{a}nsicke
2004, private communication). To constrain $\dot{m}$ and $T_{\rm
eff,0}$ we calculate $\chi^2$ using the eight temperature measurements
of the cooling over a grid of our numerical models, spaced in intervals of
$0.05\ {\rm g\ cm^{-2}\ s^{-1}}$ in $\dot{m}$
and $500\ {\rm K}$ in $T_{\rm eff,0}$.  All models assume $\tdn=52\ {\rm days}$
with the accretion rate constant during this time. In Figure \ref{fig:contour}
we show contours of constant $\chi^2$, which favors
$\dot{m}=0.30\ {\rm g\ cm^{-2}\ s^{-1}}$ and $T_{\rm eff,0}=14,500\ {\rm K}$
(shown by a star).

\begin{figure}[tbp]
\plotone{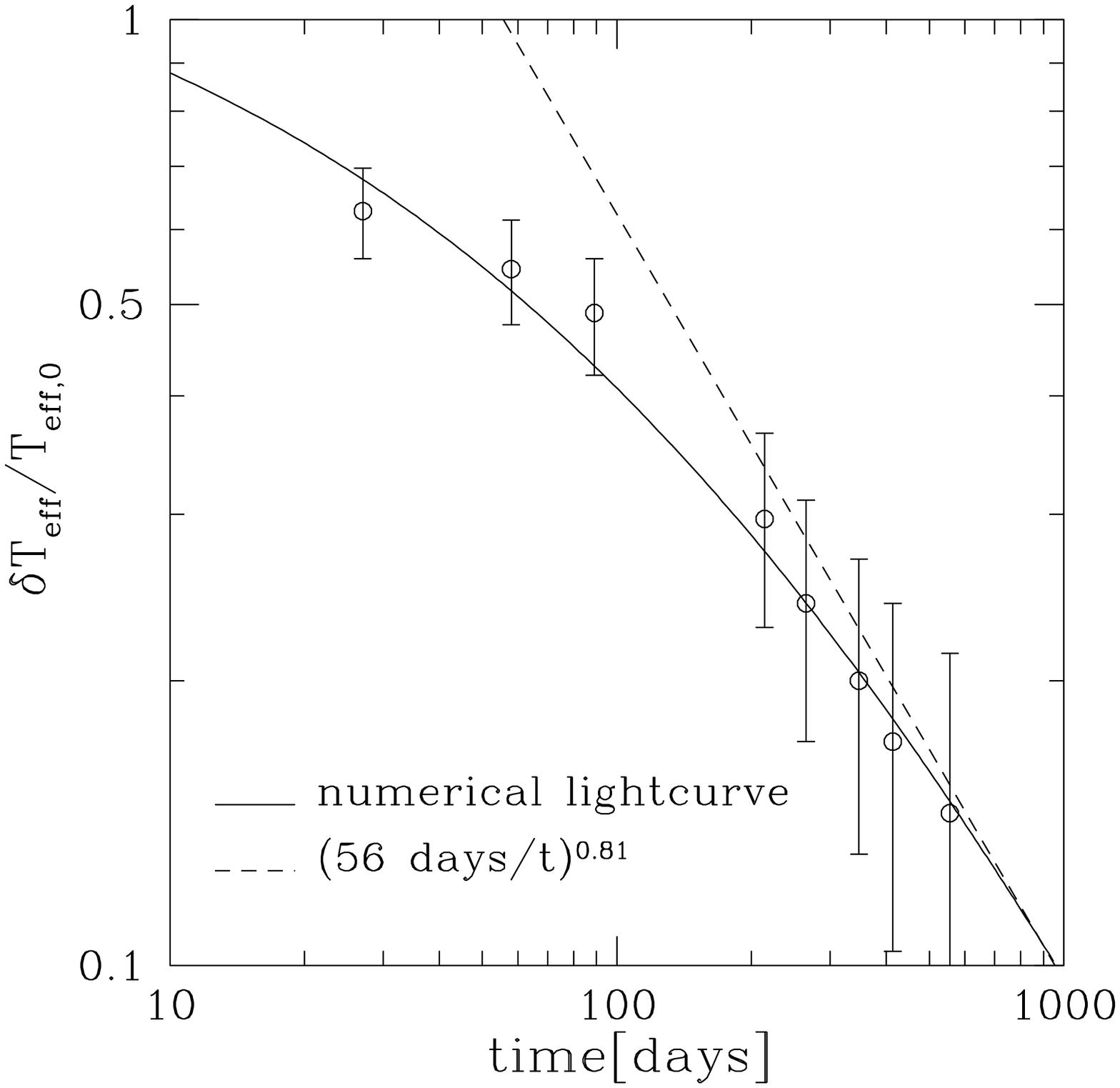}
\caption{The fractional temperature during the 2001 July outburst of
WZ Sge in comparison the cooling of a compressionally
heated envelope. The circles show temperature ``$T_b$'' from
\citet{god04}, with the error bars of $1000\ {\rm K}$.
The solid curve is our numerical calculation with
$T_{\rm eff,0}=14,500\ {\rm K}$ and
constant $\dot{m}=0.30\ {\rm g\ cm^{-2}\ s^{-1}}$ for
52 days, giving $\Delta y=1.3\times10^6\ {\rm cm}^{-2}$.
This corresponds to an average accretion rate of
$\langle\dot{M}\rangle\approx10^{-8}M_\odot\ {\rm yr}^{-1}$ for
an $M=1.1M_\odot$, $R=5\times10^{8}\ {\rm cm}$ WD. The analytic formula
given in equation (\ref{eq:fit}) for $t_{\rm late}=56\ {\rm days}$ is
shown by the dashed line.}
\label{fig:wzsge}
\end{figure}

  In Figure \ref{fig:wzsge} we plot the numerical lightcurve of this
favored model along with the corresponding analytic fit using
$t_{\rm late}=56\ {\rm days}$. At late times the measurement errors are
of order the temperature perturbations, which at face value should
make comparisons to our numerical models difficult. Fortunately, the data
show a clear trend that closely follows the general features of our
numerical lightcurve, which strongly suggests that the cooling is due to
compressional heating. It has a shallow slope at early times
and then becomes a power law at late times with
$\delta T_{\rm eff}/T_{\rm eff,0}\propto t^{-0.81}$ as
we predict. Our fit implies an accreted column
$\Delta y =1.3\times10^{6}\ {\rm g\ cm}^{-2} g_8^{-2/21}$, corresponding
to an average accretion rate during the outburst
$\langle\dot{M}\rangle\approx10^{-8}\ M_\odot\ {\rm yr}^{-1}$
(assuming a WD radius of $5\times10^8\ {\rm cm}$, or mass $M=1.1\ M_\odot$),
in reasonable agreement
with the $\dot{M}$'s estimated by \citet{lon03} of $(1-3)\times10^{-9}\
M_\odot\ {\rm yr}^{-1}$ (measured when the outburst was especially bright).
We therefore favor a massive WD (small radius) to get an $\dot{M}$ closer to
these measurements. Such a mass estimate is consistent with the value
estimated by \citet{pat02} of $1.0\pm0.2M_\odot$.
\citet{god04} also find $T_{\rm eff,0}=14,500\ {\rm K}$ and $\Delta
y\approx1.3\times10^6\ {\rm g\ cm}^{-2}$ (estimated from their Fig. 2)
are needed to explain the cooling curve. Note that they include a 
prescription for boundary layer heating, while we have ignored this effect.
The agreement between our results and \citet{god04} indicates boundary
layer heating is not important in the late time lightcurve (also see Appendix).

%  These mass estimates for the WD, along with recent estimates of $q$ for
% the binary ($0.057\pm0.017$, Steeghs et al. 2001; $0.045\pm0.020$,
% Patterson 2001), constrain the secondary to have a mass of
% $\sim0.05-0.07M_\odot$. This implies that this is not a main sequence
% object, and most likely a brown dwarf. This is consistent with measurements
% of WZ Sge in quiescence, which shows no spectral or photometric features
% of a secondary \citep{cia98,dhi00,lit00} so that the secondary has
% $M_K>12.2$ as expected for a brown dwarf
% \citep[as was previously noted by][]{pat02}.

  If we fit the last five out of eight temperatures measurements using the
analytic formula from equation (\ref{eq:fit}) instead of our numerical calculations,
we find the best fit for
$t_{\rm late}=39\ {\rm days}$ with $T_{\rm eff,0} = 15,000\ {\rm K}$.
For this fit we assume a power law index of 0.81 because the current data
does not allow us to treat it as a free parameter. This index value
is fairly robust since both our work here and similar calculations using
OPAL opacities give similar results.
Using equation (\ref{eq:deltay}), this fit implies an accreted column
$\Delta y =1.1\times10^{6}\ {\rm g\ cm}^{-2} g_8^{-2/21}$, within $20\%$ of
the full numerical lightcurve value. This shows the
usefulness of our fitting formula for making quick estimates of the outburst
properties. In general, equation (\ref{eq:fit}) will overpredict
$T_{\rm eff,0}$ because the lightcurve is merely approaches the power
law asymptotically (as shown in Figure \ref{fig:wzsge}), so that its predicted
columns are {\it lower limits}.

% -----------------------------------------------------------
% -----------------------------------------------------------

\section{Discussion and Conclusions}

  We presented an investigation into the cooling of a WD after a DN
outburst, focusing on late times when the luminosity from the WD is
dominated by cooling of adiabatically compressed deep layers. We
find that the lightcurve is a power law, {\it not} in $T_{\rm eff}(t)$,
but instead in $\delta T_{\rm eff}(t)/T_{\rm eff,0}$. This results in a
fitting function, equation (\ref{eq:fit}), that can be used to constrain
$T_{\rm eff,0}$ and the accreted column, $\Delta y$ (eq. [\ref{eq:deltay}]),
from the observations. The sensitivity to $T_{\rm eff,0}$ may be useful
if outbursts are too closely spaced for the WD to ever cool to
the quiescent temperature. Combining this with the work of \citet{tb03},
would allow one to get a better handle on the long term accretion history
over the CV's lifetime. Previous studies that do this require careful
measurements of $T_{\rm eff,0}$, but as our work shows, compressional
heating during DN outbursts can have a long lasting impact on the flux
leaving a WD. In our example models we find a $10\%$ deviation from the
quiescent flux 1000 days after the outburst has ended!

  Our study highlights the importance of multiple measurements for a given
outburst, over many epochs so as to tightly constrain the late time
cooling. Many DN outbursts have only two measurements of their cooling
lightcurve, so we cannot extend our analysis to these other systems.
We made comparisons with WZ Sge because it is the best opportunity
to test whether compressional heating is occurring and to see how well
$\Delta y$ and $T_{\rm eff,0}$ can be constrained, but even in this case
the measurement errors are frustratingly large. It is encouraging that our
work compares favorably to the outburst from this object, and we look forward
to new, more detailed measurements that will test the effectiveness of our
results.
Multiple measurements would be a powerful tool in conjunction with
our fitting function to study the properties of DN outbursts, especially
in comparing multiple outbursts from the same system using
equation (\ref{eq:mass}).

  The scalings and arguments we use may help to understand the
heating of stellar surfaces in other environments. Such effects
are observable whenever $\tth(\Delta y)$ is much longer than the
characteristic viscous timescale in the outer accretion disk,
and symbiotic binaries may be the most interesting case among such
systems. Symbiotic binaries exhibit a variety of outburst types, one
of which is commonly referred to as a ``classical symbiotic outburst''
(see Sokoloski 2004 for a recent review). In these outbursts the optical
brightness increases by one to a few magnitudes over weeks or months
and then decays over a timescale of months to years. The cause of
classical symbiotic outbursts remains a mystery, but such timescales
are characteristic of the DN outbursts we study here. The application of
our work to these systems would test whether these outbursts
are also accretion events.
Symbiotics have a number of subtleties that must be correctly
incorporated before they can be modeled, since they accrete at
$\dot{M}\sim10^{-9}-10^{-6}M_\odot\ {\rm yr}^{-1}$,
considerably higher rates than the DN systems. Consequently, there
is a hotter boundary layer and a thicker disk, both of which increase
the influence of surface heating (see Appendix).
Another difference is that the temperature inversion
we quickly mention in \S \ref{sec:heating} is more pronounced when
this much mass is accreted, but most likely it will still have a negligible
effect on the late time cooling. The $T_{\rm eff,0}$ is higher for
symbiotics because of a larger time-averaged accretion rate and
possible steady nuclear burning \citep{ss94}, and this helps to
minimize any temperature inversion.

  We thank Boris G\"{a}nsicke for providing comments on a previous draft
of this paper and Danny Steeghs and Dean Townsley for thoughtful
discussions. We also thank the anonymous referee for helpful comments and
criticisms. This work was supported by the National Science Foundation
under grants PHY99-07949 and AST02-05956, and by the Joint Institute for
Nuclear Astrophysics through NSF grant PHY02-16783. Phil Arras is an NSF
AAPF fellow.

% -----------------------------------------------------------
% -----------------------------------------------------------

\appendix

\section{Surface Heating}

\label{sec:bl}

  To this point we have neglected the effects of surface and shear heating.
We now include the effect of a hot layer heating the WD surface, during
the outburst and the subsequent cooling. Numerical runs with and without
compressional heating are done for comparison, and we present analytical
results for the case of surface heating alone. The result of our analysis
is that a surface temperature $\gtrsim10^6\ {\rm K}$ is needed for an
appreciable deviation from the analytic formula in equation (\ref{eq:fit}).
We conclude that surface heating has a negligible effect on late time cooling
because the boundary layer is not sufficiently hot. These results also
apply to heating due to the shearing of the surface accretion flow against
the WD surface, because the shearing takes place at a column much less
than those of interest for compressional heating \citep{pb04}. Another
mechanism that heats the surface is the advection of hot boundary layer
fluid in the accretion flow as considered by \citet{pop97}. Such heating
can only occur down to a column of $\Delta y$, and furthermore, this fluid
quickly cools in the shallow surface layers near the boundary layer
(because $t_{\rm th}$ is
so short there), long before it can even reach this depth. Advection is
therefore washed out by the surface heating at the depths of interest,
which are many orders of magnitude deeper.

  We follow \citet{pri88} and model boundary layer irradiation by
enforcing a surface temperature $\Tbl$ during the outburst at
the outermost grid point.  However, surface heating from irradiation only
occurs at low latitudes underneath the boundary layer.  This has been
shown both observationally \citep{mau04} and theoretically \citep{pb04}
to cover a small fraction of the WD surface area, $\epsilon_{\rm bl} \sim
0.01-0.1$, while compressional heating occurs over the whole star. Hence,
we assume that the lightcurve is generated by a fraction $\epsilon_{\rm bl}$
of the star undergoing surface and compressional heating, and a fraction
$1-\epsilon_{\rm bl}$ with compressional heating alone. In more detail,
if $T_{\rm eff,1}$ includes only compressional heating and 
$T_{\rm eff,2}$ includes compressional and surface heating, we make a 
combined lightcurve as $T_{\rm eff}=[(1-\epsilon_{\rm bl})T_{\rm eff,1}^4
+ \epsilon_{\rm bl}T_{\rm eff,2}^4]^{1/4}$. Other possible heating effects,
such as irradiation through the radial flux in the disk \citep{reg83}
or direct irradiation from the inner region of the accretion disk, will be of
the same order as the $\epsilon_{\rm bl}$ we consider here, and therefore do
not need to be considered separately.

\begin{figure}[tbp]
\epsscale{0.7}
\plotone{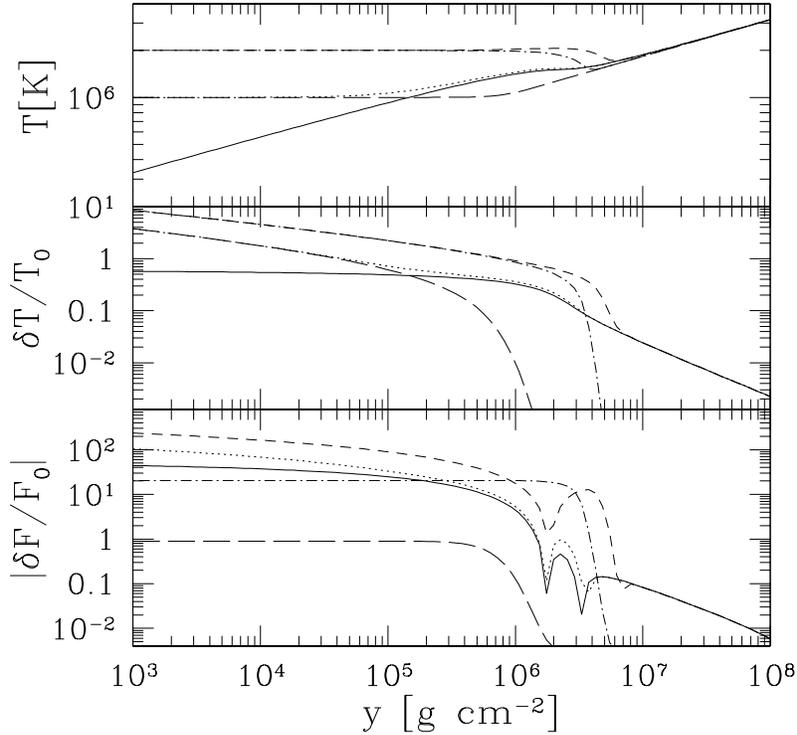}
\caption{Temperature, fractional temperature perturbation, and
fractional flux perturbation at the end of a 52 day outburst in which
the star is heated by a boundary layer and/or compressional heating.
The lines are $(\dot{m}[{\rm g\ cm^{-2}\ s^{-1}}],T_{\rm bl}[{\rm
K}])= (0.30,0.0)$ (solid), $(0.30,1.0\times 10^6)$ (dotted),
$(0.30,2.0\times 10^6)$ (short dash), $(0.0,1.0\times 10^6)$ (
long dash), and $(0.0,2.0\times 10^6)$ (dot-dashed), all starting with a
quiescent background model with $T_{\rm eff,0}=14,500\ {\rm K}$.}
\label{fig:blheatingprofile}
\end{figure}

  Figure \ref{fig:blheatingprofile} shows the temperature, temperature
perturbation, and flux perturbation underneath the boundary layer at
the end of the outburst, just before cooling begins. As $T_{\rm bl}$ is
increased, the temperature profile is affected to larger depths. However,
runs including both compressional and surface heating always show the
$\delta T/T_0 \propto y^{-1}$ scaling at large depths, so we conclude:
{\it surface heating can only propagate in a finite distance whereas
compressional heating affects the material at arbitrarily large depths.}
Hence, at late times the cooling must eventually be given by the analytic
formulas in equations (\ref{eq:fit}) and (\ref{eq:deltay}). In practice,
for sufficiently high $T_{\rm bl}$, this asymptotic scaling may occur so
long after the outburst that for the entire observation period {\it both}
heating mechanisms still affect the light curve. We will come back to
this point.

\begin{figure}[tbp]
\epsscale{0.7}
\plotone{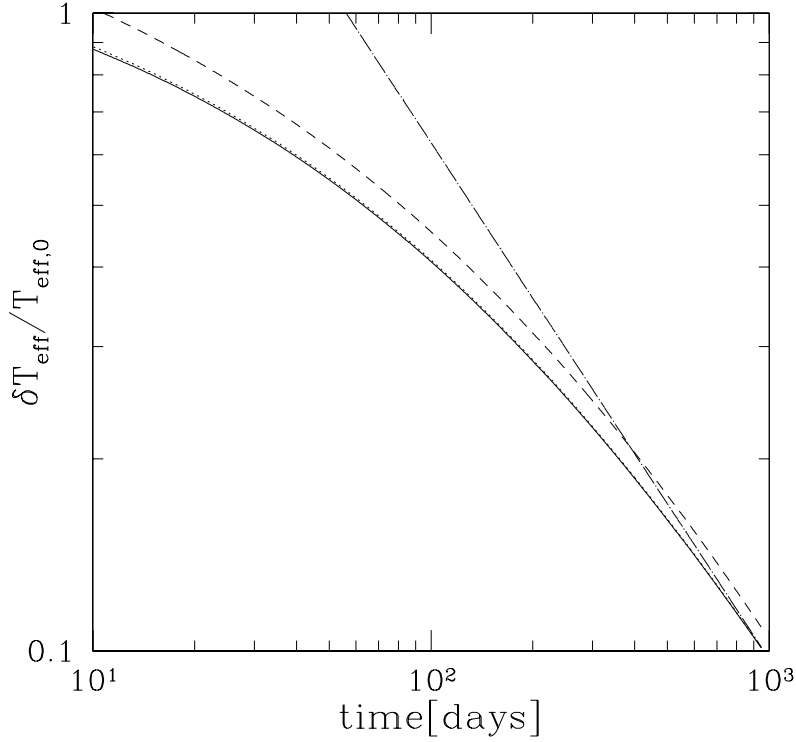}
\caption{$T_{\rm eff}$ perturbation as a function of time after end
of the outburst of duration 52 days. Lightcurves were generated for
$(\dot{m}[{\rm g\ cm^{-2}\ s^{-1}}],T_{\rm bl}[{\rm K}])= (0.30,0.0)$
, $(0.30,1.0\times 10^6)$, $(0.30,2.0\times 10^6)$.
These lightcurves were then weighted by the fraction ($10\%$) of
surface area undergoing heating, as described in the text.
The curves represent compressional heating alone (solid line),
compressional and surface heating for
$T_{\rm bl}=1.0 \times 10^6\ {\rm K}$ (dotted line)
and $T_{\rm bl}=2.0 \times 10^6\ {\rm K}$ (dashed line),
all for a quiescent background model with $T_{\rm eff,0}=14,500\ {\rm K}$.
The dot long-dashed line is the analytic formula
from equations (\ref{eq:fit}) and (\ref{eq:deltay}).}
\label{fig:blcooling}
\end{figure}

  Figure \ref{fig:blcooling} shows cooling after the outburst.
An accretion rate $\dot{m}=0.30\ {\rm g\ cm^{-2}\ s^{-1}}$ was used
for all curves, as in Figure \ref{fig:blheatingprofile}. The curves with
surface heating use $10\%$ of the surface area being heated by the
boundary layer, a plausible upper limit.  The three curves represent
compressional heating alone (solid line), and compressional and
surface heating, weighted by the surface area, for
$T_{\rm bl}=1.0 \times 10^6\ {\rm K}$ (dotted line) and
$T_{\rm bl}=2.0 \times 10^6\ {\rm K}$
(dashed line), respectively.  Surface heating is quite sensitive to
$T_{\rm bl}$, as the curve with $T_{\rm bl}=1.0\times 10^6\ {\rm K}$
is identical to that with no boundary layer heating, while
the curve with $T_{\rm bl}=2.0\times 10^6\ {\rm K}$ shows $1\%$
differences in $T_{\rm eff}$ at 1000 days.

  For the fiducial accreted column used in this appendix, which is the
column inferred for the July 2001 outburst of WZ Sge in \S
\ref{sec:wzsge}, surface heating does {\it not} affect the lightcurve
when the surface area is taken into account. As we have probably used
an overestimate of the covering fraction $\epsilon_{\rm bl}$, one
would have to increase $T_{\rm bl}$ to much larger values to compete
with compressional heating at late times.  Such high boundary layer
temperatures are inconsistent with models calculated by \citet{pb04}
for $\dot{M}\approx2\times10^{-8}M_\odot\ {\rm yr}^{-1}$, which
is still more than the average accretion rate we infer in \S 4 for during the
outburst of WZ Sge of $10^{-8}M_\odot\ {\rm yr}^{-1}$.

  We now explain the qualitative features seen in Figures
\ref{fig:blheatingprofile} and \ref{fig:blcooling}. The hot
boundary initiates a thermal wave propagating into the star. The column
depth to which this nonlinear diffusion wave propagates is
estimated from equations (\ref{eq:fluxeqn}) and 
(\ref{eq:entropy})
\be
	y_{\rm diff}(t) & \sim & \left( \frac{\alpha t}{c_p}
	T_{\rm bl}^{3+a-b} \right)^{1/(a+2)} 
	\nonumber \\ & \approx & 
	1.1\times 10^7\ {\rm g\ cm^{-2}}\ g_8^{-1/3}
	\lp \frac{T_{\rm bl}}{10^6\ {\rm K}}\rp^{17/6}
	\lp \frac{t}{1\ {\rm day}}\rp^{1/3},
\ee
where $T_{\rm bl,6}\equiv\Tbl/10^6 {\rm K}$.
At depths smaller than $y_{\rm diff}(t)$, the temperature is nearly
$T_{\rm bl}$ and the flux is approximately 
constant with the value
\be
	F & \sim & \alpha \frac{T_{\rm bl}^{4+a-b}}{y_{\rm diff}^{a+1}(t)}
	\nonumber \\ & \approx & 
	10^{14}\ {\rm erg\ cm^{-2}\ s^{-1}}\ g_8^{-1/3}
	\lp \frac{T_{\rm bl}}{10^6\ {\rm K}}\rp^{7/2}
        \lp \frac{t}{1\ {\rm day}}\rp^{-2/3}.
\ee
Given a sufficient outburst duration,
$\tdn \ga 3\ {\rm yr}\ g_8^{-1/2} F_{0,12}^{-3/2} (T_{\rm bl}/10^6\ {\rm K})^{21/4}$,
the thermal wave will
penetrate to a depth at which $T_{\rm bl}$ is equal to the background
temperature profile. This occurs at a column
\be
	y_{\rm bl} & = & \left[ (n+1) \frac{F_0}{\alpha} \right]^{-1/(a+1)}
	T_{\rm bl}^{n+1}
	\nonumber \\ & \approx & 
	1.2\ \times 10^6\ {\rm g\ cm^{-2}}\ g_8^{-1/2} F_{0,12}^{-1/2}
	\lp \frac{T_{\rm bl}}{10^6\ {\rm K}}\rp^{17/4}.
\ee
A simple estimate for when surface heating will affect the late time
cooling is to set $y_{\rm bl}$ equal to the accreted column $\Delta y$. At this point,
the temperature perturbation due to compressional heating becomes small
($\delta T/T_0 \ll 1$), and additional boundary layer heating can delay
the late time power law. We find the critical boundary layer temperature
\be
	T_{\rm bl,crit}
	&\sim& 10^6\ {\rm K}\ g_8^{2/17}F_{0,12}^{2/17}
	\lp\frac{\Delta y}{10^6\ {\rm g\ cm^{-2}}}\rp^{4/17},
	\label{eq:tblcrit}
\ee
consistent with Figure \ref{fig:blheatingprofile}. This derivation only
considers layers directly below where surface heating is occurring, so
that if $\epsilon_{\rm bl}\ll1$ then $T_{\rm bl,crit}$ must be somewhat
larger. Nevertheless, this still provides a useful lower limit for when
the boundary layer may be important. It is interesting that
this $T_{\rm bl,crit}$ is comparable to, but most likely ruled out by,
observations \citep{mau04} and theoretical modeling \citep{pn95,pb04}
of boundary layer temperatures. If $\Tbl$ is larger than this critical
value, the cooling profile does not reach our analytic result in an
observable amount of time, as shown in Figure \ref{fig:blcooling}, so
that if our fitting formula is incorrectly applied it will result in
an overprediction of $\Delta y$.

  Now consider surface heating at large depths, ignoring compressional
heating for simplicity. At depths $y \gg y_{\rm bl}$, the temperature
perturbations become small, and linear theory is again applicable.
Using the same method as in \S \ref{sec:selfsimilar}, we insert the ansatz 
$\delta T/T_0=f(\xi)$ into equations (\ref{eq:fluxeqn}) and (\ref{eq:entropy}),
where $\xi=[t_{\rm th,0}(y)/t]^{1/2}$ uses the thermal time of the background
state. We can derive a self-similar equation for the temperature perturbation
at large depths during heating, which we do not present here. As we desire
the temperature perturbations deep in the envelope, we take the
$\xi \gg 1$ limit, finding the simpler equation
\be
	\frac{d^2f}{d\xi^2}
	& \approx & - \frac{2(n+1)}{(n+2)^2} \xi \frac{df}{d\xi},
\ee
with a solution at $t=\tdn$ of
\be
	\frac{\delta T}{T_0}
	& \approx & f \propto \xi^{-1} \exp\lp-\lambda\xi^2\rp
	\nonumber \\ & \sim & 
	\left[ \frac{\tdn}{t_{\rm th,0}(y)} \right]^{1/2}
	\exp\left[ -\lambda\frac{t_{\rm th,0}(y)}{\tdn} \right],
\ee
where $\lambda=(n+1)/(n+2)^2 \approx 0.154$. Hence the temperature
perturbations ``tunneling" further than $y_{\rm bl}$ are exponentially
small, as seen in the rapid dropoffs in Figure \ref{fig:blheatingprofile}.
The small factor $\lambda \ll 1$ implies that the dropoff occurs at a
column an order of magnitude larger than one would na\"{i}vely expect.

% -----------------------------------------------------------
% -----------------------------------------------------------


\begin{thebibliography}{}
\bibitem[Bildsten(1998)]{bil98}
Bildsten, L. 1998, in The Many Faces of Neutron Stars, ed. R. Buccheri,
J. van Paradijs, \& A. Alpar (Dordrecht: Kluwer), 419
\bibitem[Cheng et al.(2000)]{che00}
Cheng, F. H., Horne, K., Marsh, T. R., Hubeny, I., \& Sion, E. M. 2000,
\apj, 542, 1064
\bibitem[G\"{a}nsicke \& Beuermann(1996)]{gb96}
G\"{a}nsicke, B. T. \& Beuermann, K. 1996, \aap, 309, L47
\bibitem[Godon, Regev, \& Shaviv(1995)]{grs95}
Godon, P., Regev, O., \& Shaviv, G. 1995, \mnras, 275, 1093
\bibitem[Godon \& Sion(2002)]{gs02}
Godon, P. \& Sion, E. M. 2002, \apj, 566, 1084
\bibitem[Godon \& Sion(2003)]{gs03}
Godon, P. \& Sion, E. M. 2003, \apj, 586, 427
\bibitem[Godon et al.(2004)]{god04}
Godon, P., Sion, E. M., Cheng, F., G\"{a}nsicke, B. T., Howell, S.,
Knigge, C., Sparks, W. M., \& Starrfield, S. 2004, \apj, 602, 336
\bibitem[Iglesias \& Rogers(1996)]{ir96}
Iglesias, C. A. \& Rogers, F. J. 1996, \apj, 464, 943
\bibitem[Ishioka et al.(2001)]{ish01}
Ishioka, R., et al. 2001, IAU Circ., 7669, 1
\bibitem[Kiplinger, Sion, \& Szkody(1991)]{kss91}
Kiplinger, A., Sion, E. M., \& Szkody, P. 1991, \apj, 366, 569
\bibitem[Long et al.(2003)]{lon03}
Long, K. S., Froning, C. S., G\"{a}nsicke, B. T., Knigge, C., Sion, E. M., \&
Szkody, P. 2003, \apj, 591, 1172
\bibitem[Long et al.(1993)]{lon93}
Long, K. S., Blair, W. P., Bowers, C. W., Davidson, A. F., Kriss, G. A.,
Sion, E. M., \& Hubeny, I. 1993, \apj, 405, 327
\bibitem[Long et al.(1994)]{lon94}
Long, K. S., Sion, E. M., Huang, M., \& Szkody, P.
1994, \apj, 424, L49
\bibitem[Long et al.(2004)]{lon04}
Long, K. S., Sion, E. M., G\"{a}nsicke, B. T., \& Szkody, P. 2004,
\apj, 602, 948
\bibitem[Mateo \& Szkody(1984)]{ms84}
Mateo, M. \& Szkody, P. 1984, \aj, 89, 863
\bibitem[Mauche(2004)]{mau04}
Mauche, C. W. 2004, \apj, 610, 422
\bibitem[Patterson et al.(2002)]{pat02}
Patterson, J. et al. 2002, \pasp, 114, 721
\bibitem[Piro \& Bildsten(2004)]{pb04}
Piro, A. L. \& Bildsten, L. 2004, \apj, 610, 977
\bibitem[Popham(1997)]{pop97}
Popham, R. 1997, \apj, 478, 734
\bibitem[Popham \& Narayan(1995)]{pn95}
Popham, R. \& Narayan, R. 1995, \apj, 442, 337
\bibitem[Pringle(1988)]{pri88}
Pringle, J. E. 1988, \mnras, 230, 587
\bibitem[Regev(1983)]{reg83}
Regev, O. 1983, \aap, 126, 146
\bibitem[Regev \& Shara(1989)]{rs89}
Regev, O. \& Shara, M. M. 1989, \apj, 340, 1006
\bibitem[Sion(1993)]{sio93}
Sion, E. M. 1993, Ann. Israel Phys. Soc., 10, 86
\bibitem[Sion(1995)]{sio95}
Sion, E. M. 1995, \apj, 438, 876
\bibitem[Sion et al.(1996)]{sio96}
Sion, E. M., Cheng, F. H., Huang, M., Hubeny, I., \& Szkody, P. 1996, \apj,
471, L41
\bibitem[Sion \& Starrfield(1994)]{ss94}
Sion E. M. \& Starrfield S. G., 1994, \apj, 421, 261
\bibitem[Szkody et al.(1998)]{szk98}
Szkody, P., Hoard, D. W., Sion, E. M., Howell, S. B., Cheng, F. H., \&
Sparks, W. M. 1998, \apj, 497, 928
\bibitem[Sokoloski(2004)]{sok04}
Sokoloski, J. L. 2004, to appear in the Journal of the American Association
of Variable Star Observers (AAVSO), astro-ph/0403004
\bibitem[Sparks et al.(1993)]{spa93}
Sparks, W. M., Sion, E. M., Starrfield, S. G., \& Austin, S. 1993, Ann.
Israel Phys. Soc., 10, 96
\bibitem[Townsley \& Bildsten(2003)]{tb03}
Townsley, D. M. \& Bildsten, L. 2003, \apj, 596, L227
\bibitem[Warner(1995)]{war95}
Warner, B. 1995, Cataclysmic Variable Stars (Cambridge: Cambridge Univ. Press)
\end{thebibliography}
\end{document}